%% file: main.tex
\documentclass[sn-nature]{sn-jnl}

\usepackage{graphicx}%
\usepackage{multirow}%
\usepackage{amsmath,amssymb,amsfonts}%
\usepackage{amsthm}%
\usepackage{mathrsfs}%
\usepackage[title]{appendix}%
\usepackage{xcolor}%
\usepackage{textcomp}%
\usepackage{manyfoot}%
\usepackage{booktabs}%
\usepackage{algorithm}%
\usepackage{algorithmicx}%
\usepackage{algpseudocode}%
\usepackage{listings}%
\usepackage{cleveref}
\usepackage{tabularx}
\newcolumntype{R}{>{\raggedleft\arraybackslash}X}
\usepackage{lmodern}
\input{math_commands}

\usepackage{colortbl}
\usepackage{subcaption} 

\usepackage{tcolorbox}
\newtcolorbox{myprompt}[2][]
{
    colframe=black,      
    colback=gray!20,     
    boxrule=1pt,         
    arc=4pt,             
    left=10pt,           
    right=10pt,
    top=10pt,            
    bottom=10pt, 
    title=#2
}

\theoremstyle{thmstyleone}%
\theoremstyle{thmstyletwo}%

\theoremstyle{thmstylethree}%

\begin{document}

\title[Combating LLM Hallucinations using Hypergraph-Driven Retrieval-Augmented Generation]{Combating LLM Hallucinations using Hypergraph-Driven Retrieval-Augmented Generation}


\author[1]{\fnm{Yifan} \sur{Feng}}\email{evanfeng97@gmail.com}

\author[2]{\fnm{Hao} \sur{Hu}}\email{huhao@stu.xjtu.edu.cn}

\author[3]{\fnm{Xingliang} \sur{Hou}}\email{HouXL@stu.xjtu.edu.cn}

\author[2]{\fnm{Shiquan} \sur{Liu}}\email{quan3759@stu.xjtu.edu.cn}

\author[4]{\fnm{Shihui} \sur{Ying}}\email{shying@shu.edu.cn}

\author[2]{\fnm{Shaoyi} \sur{Du}}\email{dushaoyi@xjtu.edu.cn}

\author[5]{\fnm{Han} \sur{Hu}}\email{hhu@bit.edu.cn}

\author[1]{\fnm{Yue} \sur{Gao}}\email{gaoyue@tsinghua.edu.cn}

\affil[1]{\orgdiv{School of Software}, \orgname{Tsinghua University}, \orgaddress{ \postcode{100871}, \state{Beijing}, \country{China}}}

\affil[2]{\orgdiv{Institute of Artificial Intelligence and Robotics}, \orgname{Xi’an Jiaotong University}, \orgaddress{\postcode{710049}, \state{Xi’an}, \country{China}}}

\affil[3]{\orgdiv{School of Software Engineering}, \orgname{Xi’an Jiaotong University}, \orgaddress{\postcode{710049}, \state{Xi’an}, \country{China}}}

\affil[4]{\orgdiv{Department of Mathematics, School of Science}, \orgname{Shanghai University}, \orgaddress{\postcode{200000}, \state{Shanghai}, \country{China}}}

\affil[5]{\orgdiv{School of Information and Electronics}, \orgname{Beijing Institute of Technology}, \orgaddress{\postcode{100871}, \state{Beijing}, \country{China}}}


\abstract{Large language models (LLMs) have transformed various sectors, including education, finance, and medicine, by enhancing content generation and decision-making processes. However, their integration into the medical field is cautious due to hallucinations, instances where generated content deviates from factual accuracy, potentially leading to adverse outcomes. To address this, we introduce Hyper-RAG, a hypergraph-driven Retrieval-Augmented Generation method that comprehensively captures both pairwise and beyond-pairwise correlations in domain-specific knowledge, thereby mitigating hallucinations. Experiments on the NeurologyCrop dataset with six prominent LLMs demonstrated that Hyper-RAG improves accuracy by an average of 12.3\% over direct LLM use and outperforms Graph RAG and Light RAG by 6.3\% and 6.0\%, respectively. Additionally, Hyper-RAG maintained stable performance with increasing query complexity, unlike existing methods which declined. Further validation across nine diverse datasets showed a 35.5\% performance improvement over Light RAG using a selection-based assessment. The lightweight variant, Hyper-RAG-Lite, achieved twice the retrieval speed and a 3.3\% performance boost compared with Light RAG. These results confirm Hyper-RAG's effectiveness in enhancing LLM reliability and reducing hallucinations, making it a robust solution for high-stakes applications like medical diagnostics.}




\maketitle

\input{0_intro}

\input{1_results}
\input{2_discussion}
\input{3_method}



\bibliography{my}

\end{document}

%% file: math_commands.tex
\usepackage{xcolor}

\usepackage{pifont}
\newcommand{\cmark}{\ding{51}} 
\newcommand{\xmark}{\ding{55}} 

\usepackage{amsmath,amsfonts,bm}

\def\eqref#1{equation~\ref{#1}}

\def\1{\bm{1}}

\def\vq{{\bm{q}}}

\def\mM{{\bm{M}}}

\DeclareMathAlphabet{\mathsfit}{\encodingdefault}{\sfdefault}{m}{sl}
\SetMathAlphabet{\mathsfit}{bold}{\encodingdefault}{\sfdefault}{bx}{n}

\def\gD{{\mathcal{D}}}
\def\gE{{\mathcal{E}}}

\def\gG{{\mathcal{G}}}

\def\gK{{\mathcal{K}}}

\def\gP{{\mathcal{P}}}

\def\gV{{\mathcal{V}}}

\def\gX{{\mathcal{X}}}



%% file: 0_intro.tex

Large language models (LLMs) have revolutionized numerous sectors through their advanced content generation capabilities. In education, they enable personalized learning pathways\cite{razafinirina2024pedagogical,naseer2024integrating}; in information retrieval, they enhance the precision and relevance of search results\cite{dagdelen2024structured,prince2024opportunities,zhao2024dense}; in finance, they improve predictive analytics and support strategic decision-making\cite{cao2022ai,nahar2024advanced}; in medicine, they assist with preliminary diagnostics and patient management\cite{ullah2024challenges,savage2024diagnostic}; and in elder care, they facilitate cognitive engagement and support for daily living activities\cite{lukkien2023toward}. Despite these advancements, the integration of LLMs within the medical domain has been relatively cautious. This hesitancy primarily stems from concerns regarding the accuracy and reliability of the generated content, which can introduce uncertainty into clinical decision-making processes and potentially lead to adverse medical outcomes\cite{quinn2021trust,hussain2022modern,hager2024evaluation}. LLMs are adept at interpreting input data and generating responses based on their training data, often exhibiting high confidence in their outputs. However, this confidence does not inherently guarantee factual correctness, resulting in discrepancies commonly referred to as LLM hallucinations\cite{huang2025survey}.

LLM hallucinations occur when the generated content diverges from established facts, colloquially termed as “bullshit.” For instance, in the diagnosis of neurological disorders, an LLM might incorrectly attribute symptoms to an unrelated condition, potentially misleading healthcare professionals\cite{quinn2021trust,hager2024evaluation,hussain2022modern}. Extensive research has been conducted to elucidate the underlying causes of these hallucinations, with findings suggesting that they likely arise from the models' training methodology, characterized by “data compression\cite{buciluǎ2006model}.” The training process typically involves self-supervised tasks that compress and reconstruct vast datasets. While LLMs can accurately reconstruct approximately 98\% of the training data, the remaining 2\% may result in significantly inaccurate or misleading responses\cite{jones2025ai}. Enhancing the models' capabilities can mitigate the frequency of hallucinations; however, the persistent "last mile" challenge continues to impede their reliable application in contexts that demand stringent adherence to factual accuracy, such as in medical practice. However, strategies aimed at enhancing the capabilities of LLMs entail substantial costs, often necessitating significant computational resources to train new models from scratch. This resource-intensive process poses scalability challenges and limits the feasibility of frequent model updates. Moreover, these enhancement strategies do not fully mitigate the loss of knowledge induced by data compression during training. As a result, even with increased model capacity, certain informational gaps and inaccuracies persist, underscoring the need for alternative approaches to preserve and integrate comprehensive knowledge without incurring prohibitive costs\cite{ke2021achieving}.

To enhance LLMs' capacity to retain and comprehend critical knowledge, thereby mitigating hallucinations, retrieval-augmented generation (RAG)\cite{prince2024opportunities,es2024ragas,xiong2024benchmarking,naiverag,miao2024integrating}, strategies have garnered extensive scholarly attention. RAG operates by constructing domain-specific knowledge repositories and employing vector-based retrieval techniques to extract pertinent prior information related to a given query. By constraining the generative process with this external knowledge, RAG enables LLMs to produce more accurate and reliable content, particularly concerning sensitive data such as numerical values or product names\cite{naiverag}. For instance, in the medical domain, the application of RAG allows LLMs to precisely identify medication names, dosages, and administration schedules\cite{xiong2024benchmarking}. In scenarios where hallucinations might lead to erroneous key information, the model's output may appear coherent and logically sound, yet critical inaccuracies can result in severe repercussions, including medical errors\cite{quinn2021trust,hager2024evaluation,hussain2022modern}. Thus, RAG serves as a crucial mechanism to ensure the fidelity of LLM-generated information in high-stakes environments.

The efficacy of RAG hinges fundamentally on the representation of domain-specific knowledge, spawning a diverse array of methodologies. The most rudimentary form of RAG\cite{naiverag} involves partitioning the raw corpus into manageable chunks and employing keyword-based retrieval to identify segments pertinent to a given query. Advancements in this domain have led to graph-based organizational strategies, exemplified by seminal approaches such as GraphRAG\cite{graphrag} and LightRAG\cite{lightrag}. GraphRAG enhances retrieval precision by extracting comprehensive knowledge graphs from the corpus and establishing hierarchical correlations among entities through clustering techniques. In contrast, LightRAG introduces a dual-layered knowledge graph architecture, comprising both local and global structures, to effectively organize and index granular details alongside overarching concepts within the original knowledge base. The quintessential attribute of these classical RAG methodologies lies in their ability to structurally encode the knowledge embedded within raw textual data, facilitating rapid retrieval of relevant prior information in response to specific inquiries. By leveraging these meticulously curated knowledge points, RAG frameworks empower LLMs to anchor their generative outputs in verified data, thereby mitigating the incidence of hallucinations and enhancing the factual integrity of the responses, as opposed to relying solely on the inherently compressed knowledge acquired during model training.

\begin{figure}[!t]
    \centering
    \includegraphics[width=\linewidth]{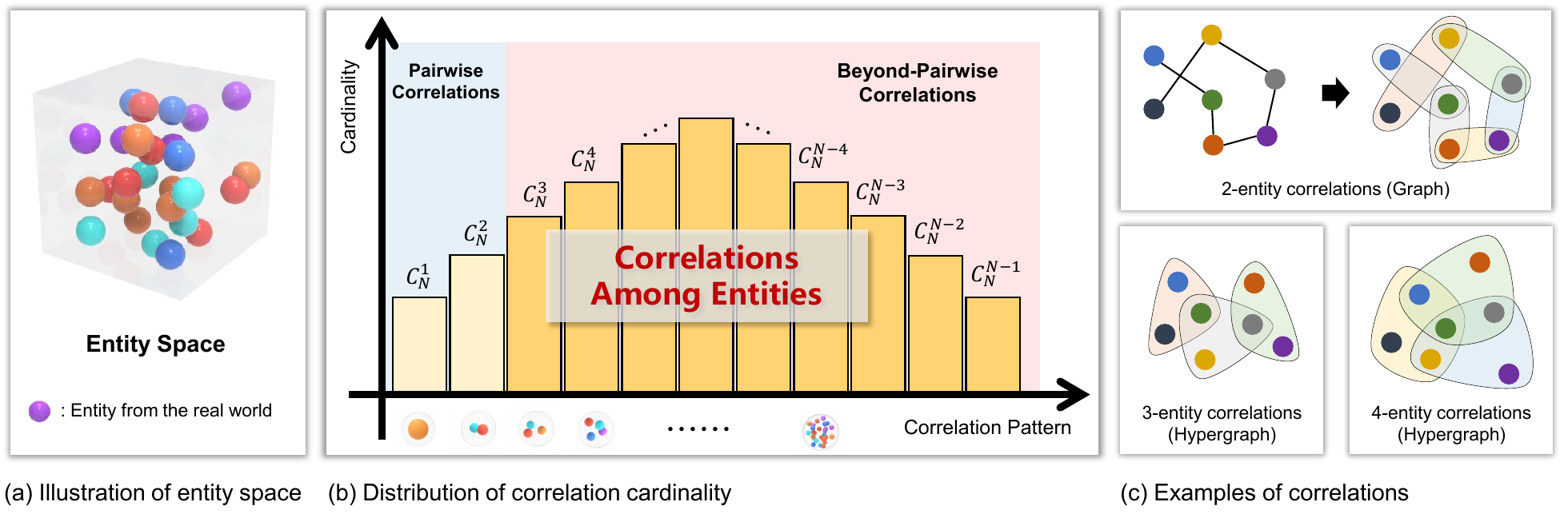}
    \caption{\textbf{Illustration of Complex Correlation Modeling in Data.} \textbf{a,} The real-world entity space, depicting the various entities present in the dataset. \textbf{b,} Potential complex correlations among these entities, including low-order correlations such as pairwise correlations or self-relations, and high-order correlations involving interactions among three or more entities. \textbf{c,} Visualization of entity correlations using circles to represent correlations between entities. The structure is modeled as a 2-uniform hypergraph, emphasizing pairwise connections. Another example illustrates correlations among three and four entities, with circles encompassing three and four entities, respectively.}
    \label{fig:hg_rep}
\end{figure}

Structuring raw corpus data can significantly enhance the efficiency of information retrieval; however, existing graph-based approaches to information architecture often result in substantial data loss\cite{srinivasan2018quantifying,santos2022knowledge}. Specifically, traditional graphs are constrained to representing pairwise correlations between entities, as illustrated in \Cref{fig:hg_rep}. In medical contexts, for example, graphs can depict binary interactions between drugs but fail to capture the complex interactions involving multiple medications simultaneously\cite{pais2024large,li2024cancergpt,singhal2023large}. Similarly, in narrative storytelling, while graphs can effectively model intricate correlations between characters, they are inadequate for representing events that involve multiple characters interacting concurrently\cite{labatut2019extraction}. These beyond-pairwise correlations are typically lost during the construction of knowledge graphs, thereby depriving LLMs of comprehensive prior information. Consequently, developing more comprehensive methods for information representation is imperative to enable LLMs to access critical knowledge and effectively mitigate the occurrence of hallucinations.

To comprehensively capture both pairwise and multi-way correlations inherent in raw data, it is imperative to adopt a modeling approach that ensures complete coverage of these correlations, thereby providing LLMs with more robust and effective prior knowledge. Hypergraphs\cite{hgnnp} emerge as a potent tool for modeling complex correlations due to their inherent flexibility. Unlike traditional graphs, where edges are limited to connecting two nodes and thus can only represent pairwise correlations, hypergraphs utilize hyperedges that can link any number of nodes, thereby facilitating the representation of multi-way correlations. As depicted in \cref{fig:hg_rep}, the correlations among points within the raw data space can be diverse, encompassing both pairwise and beyond-pairwise correlations. These varied connections collectively provide a comprehensive coverage of the possible interaction patterns within the data. Consequently, hypergraphs serve as an advanced framework for modeling inter-data correlations, enabling the complete and accurate representation of information contained within the data. This enhanced representation is crucial for empowering LLMs to access and utilize a more extensive and precise set of prior knowledge, thereby mitigating issues such as hallucinations and improving the reliability of generated outputs.

To mitigate hallucinations in LLMs, we propose a Hypergraph-Driven Retrieval-Augmented Generation method (Hyper-RAG) by incorporating hypergraph modeling into the RAG framework. Unlike existing RAG\cite{graphrag,lightrag} approaches that typically utilize traditional graph structures to represent pairwise correlations, our method leverages hypergraphs to capture the intricate and multifaceted correlations present in raw data. Specifically, low-order correlations are employed to delineate direct connections between entities, while high-order correlations and group correlations are utilized to characterize more complex interactions. The process begins with the extraction of entities from the raw dataset, which serve as nodes in the hypergraph. Subsequently, both low-order and high-order correlations between these entities are identified and integrated to construct a hypergraph-based knowledge repository. During the question-answering phase, key entities are first extracted from the input query, and relevant prior corpus information is retrieved from the knowledge base using the hypergraph structure. The inclusion of high-order correlations ensures a more comprehensive retrieval of pertinent information, thereby providing the LLM with a richer set of prior knowledge. This approach effectively compensates for the information loss resulting from the compression inherent in model training, thereby enhancing the accuracy and reliability of the generated responses.

\begin{figure}
    \centering
    \includegraphics[width=\linewidth]{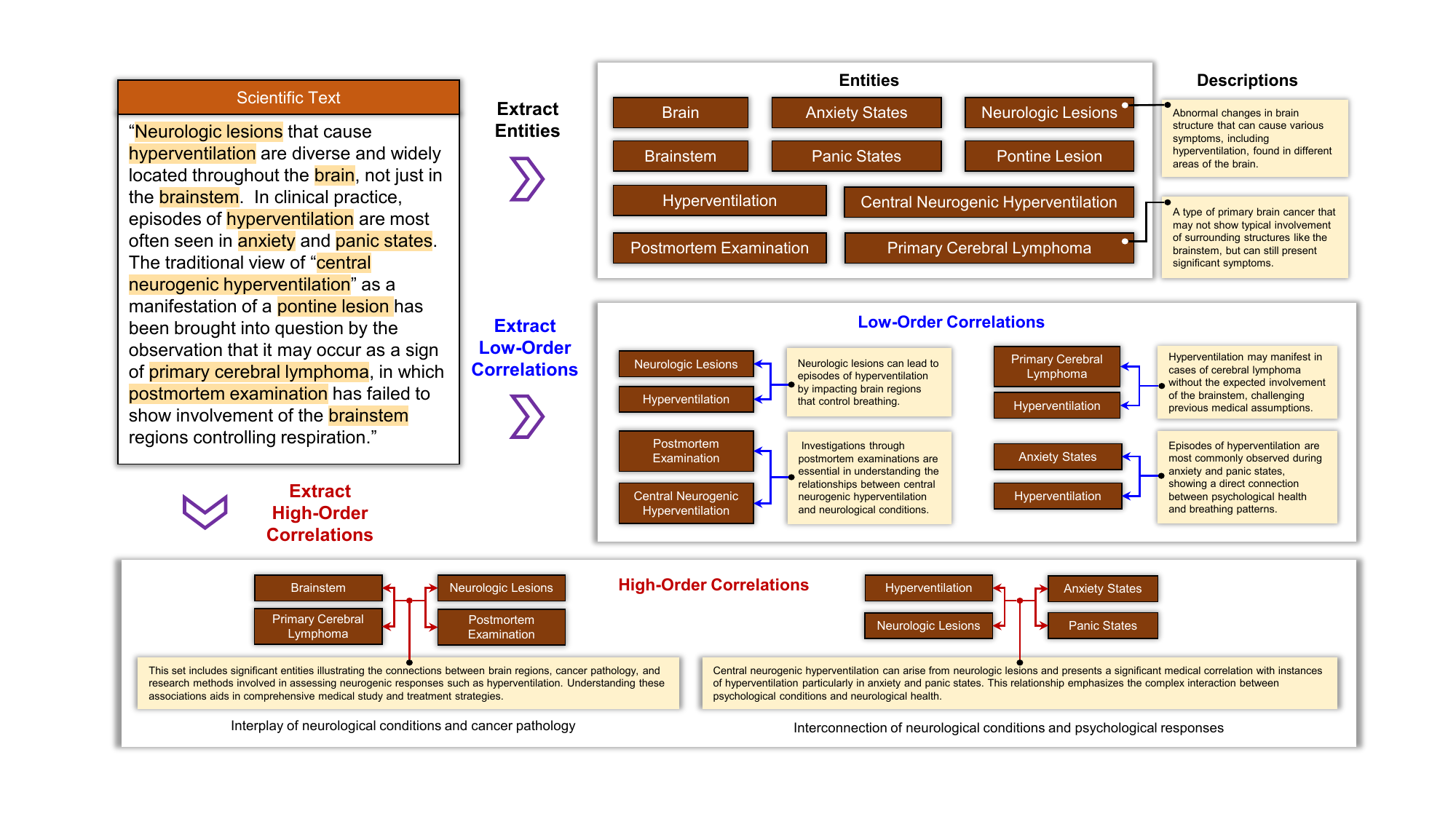}
    \caption{\textbf{Illustration of Entity and Correlation Extraction from Raw Corpus}: Dark brown boxes represent entities, blue arrows denote low-order correlations between entities, and red arrows indicate high-order correlations. Yellow boxes contain the original descriptions of the respective entities or their correlations.}
    \label{fig:extract}
\end{figure}

The core of Hyper-RAG lies in utilizing hypergraphs to achieve a comprehensive and structured representation of knowledge from raw data, thereby minimizing information loss. \Cref{fig:extract} provides an example illustrating how entities, low-order correlations, and high-order correlations are extracted from the raw corpus. For instance, consider the following excerpt from the corpus: “Neurologic lesions that cause hyperventilation are diverse and widely located throughout the brain, not just in the brainstem. In clinical practice, episodes of hyperventilation are most often seen in anxiety and panic states. The traditional view of ‘central neurogenic hyperventilation’ as a manifestation of a pontine lesion has been brought into question by the observation that it may occur as a sign of primary cerebral lymphoma, in which postmortem examination has failed to show involvement of the brainstem regions controlling respiration.” From this passage, entities such as brain, neurologic lesions, anxiety states, and hyperventilation are identified. Low-order correlations, for example, the correlation between neurologic lesions and hyperventilation, are extracted as “Neurologic lesions can lead to episodes of hyperventilation by impacting brain regions that control breathing.” Furthermore, high-order correlations involving multiple entities, such as brainstem, primary cerebral lymphoma, neurologic lesions, and postmortem examination, are also identified. These high-order correlations encompass significant entities that illustrate the connections between brain regions, cancer pathology, and research methods involved in assessing neurogenic responses like hyperventilation. This comprehensive correlation modeling facilitates a more complete knowledge structure. In contrast, if only pairwise correlations are extracted using traditional graphs, the intricate correlations among multiple entities cannot be adequately represented, leading to potential information loss. Such omissions may result in incomplete prior knowledge being available to LLMs, thereby undermining the effectiveness of RAG in mitigating hallucinations.

Hallucinations in LLMs pose significant challenges, subtly undermining the logical coherence and expressive clarity of their outputs, and overtly distorting critical nouns and factual data. These phenomena are notoriously difficult to quantify due to their diverse manifestations and the complexity of natural language understanding. Existing research and benchmark evaluations have primarily focused on assessing LLMs by posing questions with known, definitive answers to determine whether key terms and data points are accurately retrieved. While this approach provides valuable insights, it remains inherently limited and somewhat biased, as it predominantly addresses scenarios with closed-ended questions. In real-world applications, the vast majority of queries are inherently open-ended, lacking predetermined answers and often requiring nuanced, context-dependent responses of varying lengths. This discrepancy highlights a critical gap in current evaluation methodologies, which fail to capture the full spectrum of hallucination behaviors exhibited by LLMs in more complex, unconstrained environments. To bridge this gap and achieve a more comprehensive assessment of hallucinations in LLMs, we propose two novel evaluation strategies: Scoring-Based Assessment (\cref{sec:score_metric}) and Selection-Based Assessment (\cref{sec:selection_metric}). The first strategy, Scoring-Based Assessment, employs five distinct metrics to evaluate model outputs across multiple dimensions, assigning scores ranging from 0 to 100. This method provides a multifaceted evaluation framework, allowing for horizontal comparisons across various enhancement strategies and effectively quantifying the extent of hallucinations present in different models. The second strategy, Selection-Based Assessment, introduces eight metrics designed to facilitate a voting mechanism between the responses of two different models. While this approach is constrained to scenarios involving the comparison of two specific models, it enables a more granular evaluation across multiple performance aspects, offering deeper insights into the relative strengths and weaknesses of each model. By implementing these two evaluation methodologies, we aim to quantitatively measure the effectiveness of various enhancement techniques in mitigating hallucinations across different LLMs. This comprehensive assessment framework not only addresses the limitations of existing evaluation methods but also provides a robust foundation for developing strategies that enhance the reliability of LLM outputs in diverse, real-world contexts.

Intuitively, our Hyper-RAG framework achieves comprehensive coverage of prior corpus knowledge by constructing a hypergraph-driven knowledge base. This comprehensive coverage effectively guides LLMs in addressing domain-specific questions, thereby enhancing the accuracy and reliability of their responses. We conduct experiments on the NeurologyCrop dataset to evaluate the augmentation effects of Hyper-RAG on six prominent LLMs: GPT-4o Mini\cite{gpt4}, Qwen-Plus\cite{qwen}, LLaMa-3.3-70B\cite{llama}, DeepSeek-V3\cite{ds-v3}, and Doubao-1.5-Pro\cite{doubao}. The experimental results reveal that Hyper-RAG outperforms the direct application of LLMs by an average improvement of 12.3\%. Furthermore, when compared to Graph RAG and Light RAG, Hyper-RAG demonstrated additional performance gains of 6.3\% and 6.0\%, respectively. A particularly intriguing finding emerged when we manipulated the difficulty of the questions by introducing nesting—where one question is followed by another to increase complexity. As question difficulty escalated, the performance of existing LLMs and RAG-based methods exhibited significant declines. In contrast, Hyper-RAG maintain stable performance levels. Specifically, as the difficulty increased, Hyper-RAG's improvement over direct LLM usage grow from 12.7\% to 15\%. This highlights Hyper-RAG's robustness in handling more complex queries.

To further validate our approach, we extend our experiments to nine diverse corpus datasets spanning multiple domains. Across these datasets, Hyper-RAG consistently outperform the conventional graph-based method, Light RAG, achieving an average performance improvement of 35.5\% when evaluated using an alternative selection-based assessment method. Ablation studies are also conducted to assess the impact of different knowledge representations, original prior corpus, high-order correlations, and low-order correlations, on the capabilities of LLMs. The results indicated that the combined representation of high-order and low-order correlations effectively supplements information, thereby enhancing the performance of LLMs. Finally, in our performance analysis, Hyper-RAG demonstrate a balanced trade-off between speed and performance compared to graph-based methods. Notably, the lightweight variant, Hyper-RAG-Lite, which retains only the essential entity retrieval enhancements, achieved a twofold increase in retrieval speed and a 3.3\% performance improvement over Light RAG. These findings collectively substantiate the effectiveness of our Hyper-RAG method in augmenting the capabilities of LLMs and mitigating the occurrence of hallucinations.


%% file: 1_results.tex
\section{Results}


To validate the effectiveness of the proposed Hyper-RAG method, we conduct experimental evaluations on nine corpus datasets across eight domains\cite{xiong2024benchmarking}, with statistical details summarized in \cref{tab:data}. While existing LLMs demonstrate strong performance on tasks with standardized answers, their performance on open-ended responses remains modest. Therefore, in this study, we employ domain-specific open-domain question-answering (QA) tasks to assess the Hyper-RAG strategy. We design two evaluation approaches for open-ended assessments: The first involve directly scoring each model’s output across five dimensions for comparative analysis, and the second entails conducting pairwise competitions where a large language model evaluates responses from two different models based on eight metrics and casts votes accordingly. Detailed procedures can be found in \cref{sec:metric}. We select prominent LLMs, including GPT-4o Mini, Qwen-Plus, LLaMa-3.3-70B, DeepSeek-V3, and Doubao-1.5-Pro, as baselines and applied various augmentation strategies, namely, RAG, GraphRAG, LightRAG, and our proposed Hyper-RAG—to evaluate their impact on model outputs. More information on those augmentation strategies is described in \cref{sec:method}. Subsequently, we performed four sets of experiments to comprehensively assess our method.

\begin{table}[!t]
\caption{Statistical Information of the Corpus Dataset.}
\label{tab:data}
\begin{tabularx}{\linewidth}{XXRRR}
\toprule
Dataset & Domain & \#Token & \#Chunk & \#Ques \\ \midrule
NeurologyCorp & Medicine    & 1,968,716 & 1,790 & 2,173 \\
PathologyCorp & Medicine    & 905,760  & 824  & 2,530 \\
MathCrop      & Mathematics & 3,863,538 & 3,513 & 3,976 \\
AgricCorp     & Agriculture & 1,993,515 & 1,813 & 2,472 \\
FinCorp       & Finance     & 3,825,459 & 3,478 & 2,698 \\
PhysiCrop     & Physics     & 2,179,328 & 1,982 & 2,673 \\
LegalCrop     & Law         & 4,956,748 & 4,507 & 2,787 \\
ArtCrop       & Art         & 3,692,286 & 3,357 & 2,993 \\
MixCorp       & Mix         & 615,355  & 560  & 2,797 \\
\bottomrule
\end{tabularx}
\footnotetext{``\#Token'' denotes the number of tokens of the dataset, ``\#Chunk'' represents the number of chunks generated from the dataset, and ``\#Ques'' indicates the average number of tokens per "question."}
\end{table}

\subsection{Performance of Integrating with Diversity LLMs}


We first conduct experiments to evaluate the performance of the proposed Hyper-RAG when collaborating with various LLMs in order to verify whether it can effectively enhance the accuracy of LLM responses while mitigating hallucinations. Given that medical data is replete with specialized knowledge—and that even the slightest deviation in terminology can precipitate severe consequences such as misdiagnosis—we performed a comprehensive comparative study using the NeurologyCorp dataset. This dataset comprises extensive records of neuroscience knowledge and clinical practices, making it an ideal benchmark for assessing precision in a high-stakes domain.

\begin{figure}[!t]
    \centering
    \includegraphics[width=\linewidth]{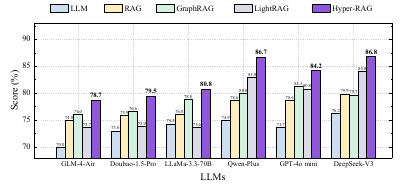}
    \caption{\textbf{Results of Integrating Hyper-RAG with Different Large Language Models.} Each LLM displayed on the x-axis represents the respective base model as indicated by its label. The other RAG methods shown are enhancements built upon these base models. The evaluation scores are calculated as the average of five scoring-based assessment metrics. The results demonstrate that Hyper-RAG consistently improves performance by an average of 12.3\% across six LLMs, highlighting its effectiveness in enhancing model capabilities through integration with LLMs.}
    \label{exp:fig:many_llms_all}
\end{figure}

To set up the experiment, the corpus are segmented into 1,968,716 chunks, and distinct prior knowledge bases are constructed for each method. For the standard Retrieval-Augmented Generation (RAG) approach, embeddings are directly extracted from each chunk and stored in a vector database to facilitate knowledge retrieval. In contrast, both Graph RAG and Light RAG establish a graph-based knowledge base: from each chunk, entity information and paired correlations are extracted and stored in a graph database, with each entity and correlation accompanied by a brief textual description. Notably, when identical vertices or paired correlations emerge across multiple chunks, their descriptive texts are merged using a large model to ensure consistency. For Hyper-RAG, we built a hypergraph knowledge base from the original neuroscience corpus. From each chunk, we extracted not only entity information and paired correlations, but also higher-order correlations that transcend pairwise relationships, with every entity and correlation supplemented by a textual description. The primary distinction between Hyper-RAG and the conventional Graph RAG lies in its inclusion of non-paired, higher-order correlations, which results in a more comprehensive and structured representation of the source data. Detailed implementation specifics are provided in \cref{sec:method}.

\begin{figure}[!t]
    \centering
    \includegraphics[width=\linewidth]{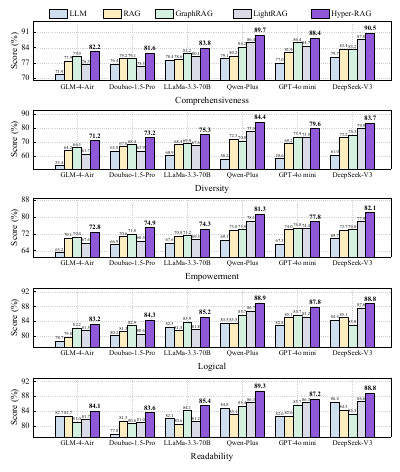}
    \caption{\textbf{Detailed results of integrating Hyper-RAG with various LLMs.}}
    \label{exp:fig:many_llms}
\end{figure}


To facilitate a robust horizontal comparison across different methods and LLMs, we adopt a scoring-based assessment that quantifies response quality using five distinct metrics (see \cref{sec:score_metric} for additional details). Prior to the experiments, 50 unique questions are randomly sampled from different chunks using the large models, ensuring that each LLM and augmentation strategy is evaluated on an identical set of queries. The experimental results are shown in \cref{exp:fig:many_llms,exp:fig:many_llms_all}, where the baseline LLMs are presented alongside the performance improvements achieved via the different augmentation approaches.


From \cref{exp:fig:many_llms,exp:fig:many_llms_all}, we have six key observations. First, compared to direct LLM responses, our proposed Hyper-RAG method yields an average improvement of 12.3\%, as shown in \cref{exp:fig:many_llms_all}. Notably, Hyper-RAG enhances performance by 15.8\% relative to Qwen-Plus and by 14.3\% in comparison to GPT-4o mini, underscoring the value of constructing a hypergraph-based prior knowledge repository for elevating the quality of LLM outputs. Second, our findings corroborate that integrating a domain-specific prior knowledge base using the RAG strategy significantly boosts response quality. Specifically, a naive RAG approach improves baseline responses by 4.9\% on average, while Graph RAG and Light RAG achieve enhancements of 6.3\% and 6.0\%, respectively. In contrast, Hyper-RAG delivers a 12.3\% enhancement over the baseline, highlighting the critical role of domain knowledge in reinforcing LLM capabilities. Third, organizing the underlying corpus with structured associative frameworks markedly bolsters RAG performance. The introduction of relational structures yields a 7.0\% improvement over unstructured methods. This increase likely stems from the fact that a well-structured representation assists in more efficient retrieval of pertinent information and promotes the diffusion of contextual cues along the relational network, thereby fostering a broader, more innovative spectrum of responses.

Fourth, both the baseline LLM and the RAG-augmented approaches demonstrate high scores in logical coherence and readability. This result reflects the extensive pre-training on large-scale corpora, which endows these models with inherent abilities to produce logically sound and accessible text regardless of the query context. Fifth, LLMs tend to score lower on metrics of comprehensiveness, diversity, and empowerment. These lower scores likely reflect intrinsic challenges in capturing nuanced domain-specific details and expressive capability. Encouragingly, the incorporation of prior corpus information via the RAG strategy results in an average improvement of 9.4\% for these metrics, thereby partially offsetting these limitations. Sixth, baseline LLMs generally exhibit modest diversity scores—typically around 60, with model such as GLM, scoring as low as 53. In contrast, implementing Graph RAG elevates diversification by 19.3\%, and our Hyper-RAG method further boosts the diversity score by 31.6\%. This substantial gain can be attributed to the integration of additional correlation information, which more effectively steers responses towards greater divergence. Moreover, the comprehensive coverage of both lower-order and higher-order correlations cultivates a richer prior knowledge base, thereby driving significant improvements across all evaluation metrics.

\subsection{Performance of Different Questioning Strategies}

\begin{figure}[!t]
    \centering
    \includegraphics[width=\linewidth]{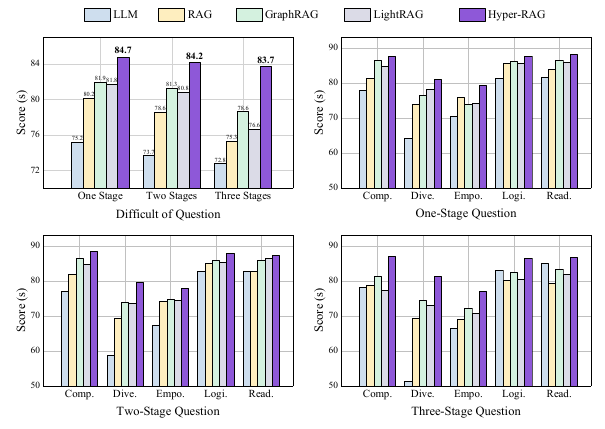}
    \caption{\textbf{Experimental results of questions with different difficulty.} The first subplot summarizes the experimental results across three different difficulties, with each score representing the average of five dimension-based assessments. The subsequent three subplots display the response quality scores across five dimensions for different methods, each targeting a specific difficulty. The x-axis displays six evaluation metrics: Comp. (Comprehensiveness), Dive. (Diversity), Empo. (Empowerment), Logi. (Logical), Read. (Readability), and Overall (the average of these five metrics).
    }
    \label{exp:fig:multi-stage}
\end{figure}

Given that our Hyper-RAG method provides a more comprehensive coverage of the knowledge embedded in the raw data, we further evaluated its capabilities by varying the difficulty of the questions. In our framework, questions are nested and asked progressively; the deeper the nesting, the greater the complexity of the task. This design is premised on the fact that a series of interdependent queries will magnify the impact of any inaccuracies in earlier responses, thereby serving as a stringent test of the LLM’s grasp of the domain knowledge. \Cref{tab:question} illustrates examples of these progressive questions, where each subsequent inquiry is formulated based on the preceding one and connected by transitional terms (indicated in bold font). We categorized the questions into three tiers according to their escalation in difficulty: single-stage, two-stage, and three-stage questions. For this experiment, GPT-4o mini is employed as the baseline LLM, and we compare several enhancement strategies, including RAG, Graph RAG, Light RAG, and our Hyper-RAG. The experimental results are presented in \cref{exp:fig:multi-stage}. 

\begin{table}
    \centering
    \caption{Examples of questions with different difficulty.}
    \label{tab:question}
    \begin{tabularx}{\linewidth}{lX}
    \toprule
        Type & Examples\\ \midrule
        One-Stage Question & What is the role of the ventrolateral preoptic nucleus in the flip-flop mechanism described for transitions between sleep and wakefulness?\\ \midrule
        Two-Stage Question & Identify the anatomical origin of the corticospinal and corticobulbar tracts, \textbf{and explain} how the identified structures contribute to the control of voluntary movement.\\ \midrule
        Three-Stage Question & How does the corticospinal system function in terms of movement control, \textbf{and specifically}, what are the roles and interconnections of the basal ganglia and the thalamus in modulating these movements, \textbf{including} the effects of lesions in these areas on movement disorders?\\ 
    \bottomrule
    \end{tabularx}
    \footnotetext{The difficulty of each question is categorized based on the number of nested layers it contains; the more nested layers, the higher the difficulty. In the examples, the bold text highlights the conjunctions that connect progressive sub-questions.}
\end{table}



\Cref{exp:fig:multi-stage} yields three key insights from our experiments. First, as evident from the first panel, Hyper-RAG consistently demonstrates stable performance improvement across various levels of question difficulty. This indicates that employing hypergraphs to represent the full spectrum of prior corpus knowledge effectively captures domain information and guides the LLM toward more accurate responses. Second, we observe that as the complexity of the questions increases, the performance of the baseline LLM gradually declines, from 75.2 to 73.7 and then to 72.8. A similar trend is observed in other RAG methods, such as Graph RAG, which reinforces the notion that more heavily nested queries place a higher demand on prior knowledge. Although all methods exhibit some degradation in performance with increasing difficulty, RAG-based approaches still manage to enhance performance relative to the original LLM. Notably, Light RAG, which omits clustering steps, loses a portion of this vital information, and its performance deteriorates more significantly as the questions become more complex.

Finally, our Hyper-RAG shows a more pronounced improvement as the difficulty increases. Specifically, relative to the baseline LLM, our method achieves incremental gains of 12.7\%, 14.3\%, and 15.0\% as the question complexity escalates, while, when contrasted with Light RAG, the improvements are 8.7\%, 9.7\%, and 5.3\%, respectively. These results substantiate the superiority of a hypergraph-based, comprehensive information representation. For straightforward queries, direct responses from an LLM or simple pairwise (i.e., low-order) correlations may suffice. However, as queries become more intricate, the availability of complex higher-order correlations becomes essential to constrain and enrich the model's outputs. \textbf{This experimental trend underscores the importance of developing hypergraph-based structural representations and retrieval methods to meet the challenges posed by increasingly complex questions.}

\subsection{Performance in Diversity Domains}
\begin{figure}
    \centering
    \includegraphics[width=0.95\linewidth]{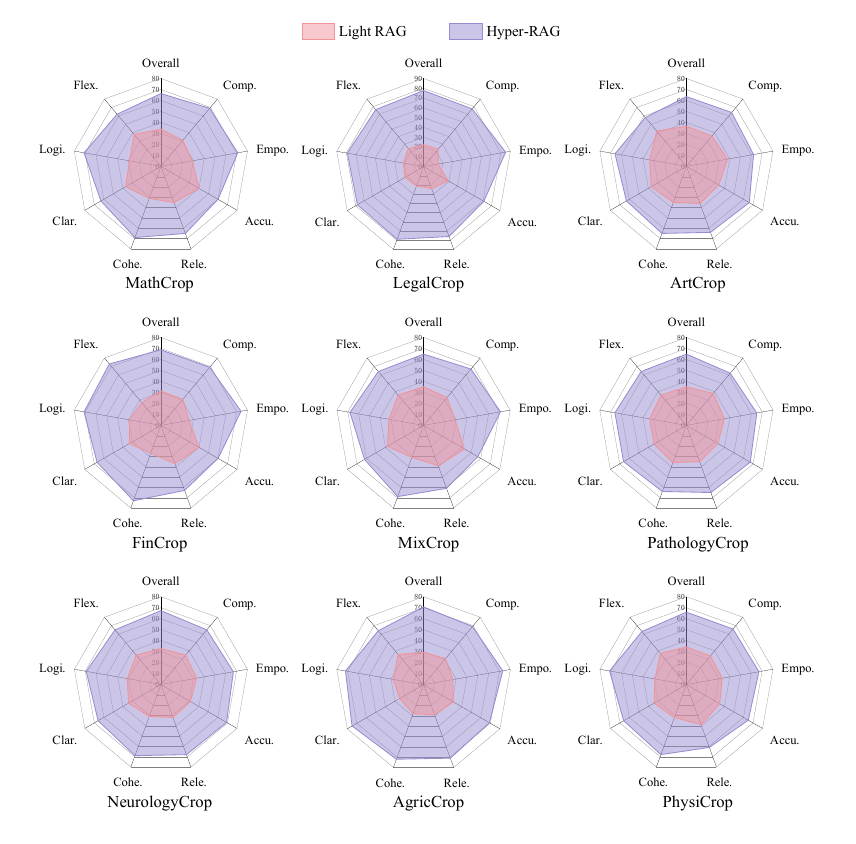}
    \caption{\textbf{Experimental results across diverse domain datasets.} This figure presents the experimental outcomes of two methods evaluated on datasets from various domains. We utilize a Selection-Based Assessment approach, employing eight distinct indicators to measure and compare the performance of the methods. The Overall score is calculated as the average of these eight evaluation metrics, providing a comprehensive assessment of each method's effectiveness. The results illustrate how the two methods perform across different domain-specific challenges, highlighting their relative strengths and areas for improvement based on the aggregated evaluation criteria.}
    \label{fig:multi-domain}
\end{figure}

To validate the adaptability of Hyper-RAG across various data domains, we further evaluate its effectiveness using nine corpora spanning eight different fields (for statistical details, see \cref{tab:data}). In this experiment, we select GPT-4o mini as the baseline architecture and compared our method against Light RAG, the latest graph-based RAG approach. Notably, we adopt a Selection-Based Assessment to comprehensively compare the performance of the graph-based and hypergraph-based RAG methods across these diverse domains. This assessment involved voting across eight distinct evaluation metrics that collectively capture the strengths and weaknesses of both approaches. For further details on the evaluation criteria, please refer to \cref{sec:selection_metric}. The experimental results are presented in \cref{fig:multi-domain}.

Based on experimental results shown in \cref{fig:multi-domain}, our Hyper-RAG method has demonstrated impressive improvements across nine datasets, with an average performance increase of 35.5\%. Specifically, the method delivers a 55.3\% improvement on LegalCrop, 41.3\% on AgricCrop, and 37.5\% on FinCrop, underscoring its effectiveness and adaptability across diverse domains. Although the enhancements in Accuracy and Relevance are relatively modest—averaging 29.8\% and 32.0\% respectively, indicating that existing low-order correlations provide a substantial baseline, Hyper-RAG exhibits notably stronger gains in Comprehensiveness and Coherence, with average improvements of 35.1\% and 39.6\%. These results can be attributed to Hyper-RAG’s unique capability to leverage both low- and high-order correlations, offering a more complete representation of the underlying data and enhancing response consistency through embedding-based retrieval from a vector database. Overall, while graph-based RAG methods may suffice for tasks primarily focused on accuracy and relevance, our Hyper-RAG method shows significant promise for more complex tasks that demand a broader and deeper domain understanding, paving the way for extensive future applications.

\subsection{Experiments of Different Knowledge Representations}
In this paper, we claim that using hypergraphs to structurally extract information from raw corpus data can more completely represent the inherent structure of the data. In this subsection, we perform an ablation study on our data organization methods based on three types of prior knowledge representations: $\gD$, $\gE_{\text{low}}$, and $\gE_{\text{high}}$. Here, $\gD$ refers to directly splitting the raw corpus into chunks and using the embedding representation of each chunk for data retrieval; $\gE_{\text{low}}$ captures pairwise knowledge correlations to construct a knowledge graph; and $\gE_{\text{high}}$ extracts non-pairwise higher-order correlations to build a knowledge hypergraph. Based on these three fundamental representations, we can construct eight different types of knowledge representations, as shown in \cref{exp:tab:org}. Evidently, directly using the LLM involves no additional data organization; using only $\gD$ corresponds to a simple RAG; combining $\gD$ with $\gE_{\text{low}}$ creates a variant similar to Graph RAG (with clustering removed from the original Graph RAG for a fair comparison); and employing all three yields our proposed Hyper-RAG. We employ GPT-4o mini as the foundational LLM and evaluated the various prior knowledge representation strategies using a scoring-based assessment (\cref{sec:selection_metric}). The experimental results are presented in \cref{exp:tab:org}.

\begin{table}[]
\caption{\textbf{Results of different knowledge representation strategies.} 
}
\label{exp:tab:org}
\begin{tabular}{cp{0.12cm}<{\centering}p{0.2cm}<{\centering}p{0.27cm}<{\centering}ccccccc}
\toprule
 Method&$\gD$ & $\gE_\text{low}$ & $\gE_\text{high}$ & Comp. & Dive. & Empo. & Logi. & Read. & Overall & Rank \\ \midrule
 LLM&\xmark & \xmark & \xmark & 77.00 & 58.60 & 67.26 & 82.80 & 82.60 & 73.65 & 8 \\
 -&\xmark & \cmark & \xmark & 83.40 & 74.96 & 74.72 & 86.06 & 86.24 & 81.08 & 6 \\
 -&\xmark & \xmark & \cmark & 84.40 & 75.00 & 75.68 & 86.46 & 86.22 & 81.55 & 5 \\
 -&\cellcolor{gray!15}\xmark & \cellcolor{gray!15}\cmark & \cellcolor{gray!15}\cmark & \cellcolor{gray!15}85.90 & \cellcolor{gray!15}78.34 & \cellcolor{gray!15}77.14 & \cellcolor{gray!15}87.02 & \cellcolor{gray!15}86.70 & \cellcolor{gray!15}83.02 & \cellcolor{gray!15}3 \\ \midrule
 RAG &\cmark & \xmark & \xmark & 81.90 & 69.24 & 74.00 & 85.06 & 82.62 & 78.56 & 7 \\
 -&\cmark & \cmark & \xmark & 85.80 & 77.20 & 76.56 & 86.58 & 86.84 & 82.60 & 4 \\
 -&\cellcolor{gray!25}\cmark & \cellcolor{gray!25}\xmark & \cellcolor{gray!25}\cmark & \cellcolor{gray!25}88.26 & \cellcolor{gray!25}78.80 & \cellcolor{gray!25}77.52 & \cellcolor{gray!25}87.34 & \cellcolor{gray!25}87.04 & \cellcolor{gray!25}83.79 & \cellcolor{gray!25}2 \\
 Hyper-RAG &\cellcolor{gray!40}\cmark & \cellcolor{gray!40}\cmark & \cellcolor{gray!40}\cmark & \cellcolor{gray!40}\textbf{88.40} & \cellcolor{gray!40}\textbf{79.60} & \cellcolor{gray!40}\textbf{77.84} & \cellcolor{gray!40}\textbf{87.76} & \cellcolor{gray!40}\textbf{87.22} & \cellcolor{gray!40}\textbf{84.16} & \cellcolor{gray!40}1 \\ \bottomrule
\end{tabular}
\footnotetext{This table presents the experimental results of various knowledge representation methods using the GPT-4o mini as the base model, conducted on the NeurologyCrop dataset. Here, $\gD$ represents the domain-specific prior corpus. $\gE_\text{low}$ denotes the low-level associative information extracted from $\gD$, and $\gE_\text{high}$ represents the high-level associative information extracted from $\gD$. The symbols \cmark and \xmark  indicate whether the respective knowledge is utilized to enhance the LLMs. 
}
\end{table}

From the experimental results presented in \cref{exp:tab:org}, we draw the following four key observations. First, employing the comprehensive data representation strategy, referred to as Hyper-RAG, yields the highest performance with a score of 84.16. This superiority is attributed to the holistic organization of data, which effectively imparts prior knowledge to the large language model. Second, we observe that augmenting the model with knowledge representations significantly enhances performance compared to the baseline without such augmentation, as illustrated in the first row. Specifically, the use of any single knowledge representation method results in an improvement of at least 4.9\%. This underscores the efficacy of knowledge augmentation strategies in enhancing the model's ability to respond accurately within domain-specific contexts.
Third, our findings indicate that utilizing the original corpus as supplementary information leads to better performance than relying solely on descriptions of entities and their correlations. The latter approach may introduce errors or hallucinations due to summarization by the large model, thereby negatively impacting the augmentation effectiveness. Lastly, when enhancing the model with a single type of correlation, high-order correlations outperform low-order ones. High-order correlations encompass more extensive information and cover a broader spectrum of knowledge within the correlation representation space, as depicted in \cref{fig:hg_rep}. In our current experiments, approximately 4,000 high-order and 13,000 low-order correlations were extracted from the prior corpus. Remarkably, the use of only high-order correlations resulted in superior performance, demonstrating a more effective enhancement of the large model. This indicates that a relatively smaller set of high-order correlations can encapsulate more substantial knowledge, thereby offering a promising new direction for the future development of RAG techniques.

\subsection{Efficiency Analysis}

\begin{figure}[]
  \subcaptionbox{Comparison of Retrieval Time\label{exp:fig:speed_time}}[0.5\textwidth][c]{
    \includegraphics[width=0.5\textwidth]{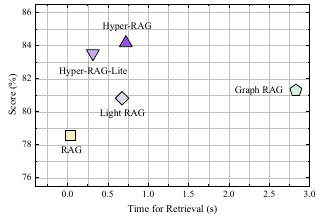}
  }
  \hfill 
  \subcaptionbox{Comparison of Performance\label{exp:fig:speed_perf}}[0.5\textwidth][c]{
    \includegraphics[width=0.5\textwidth]{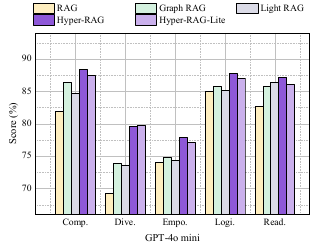}
  }
  \caption{\textbf{Efficiency Comparison of Different Augmentation Methods.} \textbf{a,} Comparison of performance and time. The performance is obtained by scoring-based assessment on the NeurologyCrop dataset, where each method's performance is the average of five indicators. The average retrieval time for RAG, Graph RAG, Light RAG, Hyper-RAG, and Hyper-RAG-Lite are 0.033s, 2.83s, 0.676s, 0.723s, and 0.315s respectively. \textbf{b,} Specific scores of the five indicators on the NeurologyCrop dataset.}
  \label{exp:fig:speed}
\end{figure}

We further conduct an efficiency analysis of the proposed method. Utilizing GPT-4o mini as the base model, we perform efficiency experiments on the NeurologyCrop dataset, comparing our Hyper-RAG approach with the fundamental RAG, Graph RAG, and Light RAG methods. To ensure a fair comparison unaffected by network latency, we exclusively evaluate the time required for local retrieval from the database to acquire relevant knowledge and the construction of the prior knowledge prompt. For the standard RAG, this primarily involves the direct retrieval time of chunk embeddings. In contrast, Graph RAG, Light RAG, and our Hyper-RAG method encompass both the retrieval time from node and correlation vector databases and the time required for a single layer of graph or hypergraph information diffusion. Since the retrieval time is influenced by both the specific questions and the methods employed, we calculate the average response time for each method by posing 50 questions from the dataset. Consistent with our selection metric outlined in \cref{sec:metric}, we employ a scoring-based assessment as the evaluation criterion. Additionally, to accommodate practical applications, we develop a lightweight variant of Hyper-RAG, termed Hyper-RAG-Lite, which preserves the essential enhancements for entity retrieval. The experimental results, presented in \cref{exp:fig:speed}, demonstrate the comparative efficiency of each method.

\Cref{exp:fig:speed_time} presents a comparative analysis of performance versus time, where points closer to the top-left corner indicate faster speeds and superior performance. From the figure, we derive the following four key observations: Firstly, we observe that both the proposed Hyper-RAG and Hyper-RAG-Lite are positioned near the top-left corner of the plot, indicating that these methods excel in both speed and performance. This demonstrates the efficacy of our approach in maintaining high efficiency without compromising answer quality. Secondly, we note that RAG is situated at the far left of the plot. This positioning is attributed to its sole reliance on document corpus retrieval without incorporating structural diffusion, which confers an efficiency advantage. However, this method's performance significantly lags behind other structure-based enhanced methods, highlighting a trade-off between speed and accuracy. Thirdly, Graph RAG achieves a performance level that is second only to our Hyper-RAG method, yet it incurs considerable time delays. The primary reason for this sluggishness is the necessity of retrieving community information in addition to node information retrieval and diffusion. Community information is derived through hierarchical clustering of nodes and lacks indexing via vector databases, necessitating layer-by-layer matching and retrieval, thereby slowing down the process. Nevertheless, the inclusion of community information, which embodies high-order correlations, effectively complements pairwise graph correlations, thereby enhancing performance. Additionally, we observe that Light RAG omits the retrieval of community information, resulting in a reduction in performance. However, this omission leads to a substantial increase in processing speed, as the computational overhead associated with managing high-order correlations is eliminated. 
Lastly, our Hyper-RAG method exhibits performance comparable to Light RAG while maintaining superior speed. Both methods essentially rely on prompt-based correlation extraction and indexing through vector databases and graph/hypergraph databases. However, Hyper-RAG differentiates itself by extracting both low-order and high-order correlations via prompts and utilizing a hypergraph database for indexing, thereby achieving similar efficiency levels. Crucially, Hyper-RAG compensates for the information loss inherent in Light RAG by incorporating additional high-order correlations, resulting in enhanced performance. It is noteworthy that the Hyper-RAG-Lite variant, although retaining only entity information retrieval, still implements diffusion through high-order correlations. This ensures that Hyper-RAG-Lite introduces additional high-order information, thereby achieving performance improvements over both Light RAG and Graph RAG.

%% file: 2_discussion.tex
\section{Discussion}






In our study, we integrate Hyper-RAG with six widely used LLMs, demonstrating a significant enhancement in performance. On average, Hyper-RAG improves the models' accuracy by 12.3\% compared to their direct application without retrieval augmentation. When juxtaposed with the conventional Graph RAG approach, our method yielded an additional 5.3\% improvement. This superior performance can be attributed to Hyper-RAG's comprehensive coverage of domain-specific knowledge. By modeling both low-order (pairwise) and high-order (beyond-pairwise) correlations within the data, Hyper-RAG facilitates a more complete and structured representation of domain knowledge, thereby reducing information loss and enhancing the quality of the generated responses.

We also evaluate the robustness of Hyper-RAG by increasing the complexity of the questions through added nesting. The experimental results reveal that existing methods, including RAG and Graph-RAG, experienced a noticeable decline in performance under these more challenging conditions. In stark contrast, Hyper-RAG maintains its performance levels, underscoring the pivotal role of high-order correlations in enabling LLMs to handle complex queries effectively. This finding suggests that current LLMs possess untapped potential for improvement in complex question-answering scenarios and that the incorporation of high-order relational modeling can significantly bolster their ability to provide accurate and reliable responses.

Our investigations into knowledge representation highlight that the impact of prior knowledge on model performance varies across different scenarios. In simpler contexts, leveraging low-order correlations alongside the original prior corpus suffices to cover the necessary information. However, in more intricate scenarios, the inclusion of high-order correlations becomes imperative to enhance the accuracy of the model’s responses. This adaptability in knowledge representation allows for the selection of appropriate prior knowledge based on the complexity of the task at hand, thereby optimizing the model’s performance across diverse application domains.

Despite its advantages, Hyper-RAG presents certain limitations. The construction of the knowledge base necessitates the extraction of high-order correlations, which introduces additional steps into the knowledge base development process. Nonetheless, the number of high-order correlations is considerably smaller compared to low-order ones, mitigating the overall impact. Moreover, these extraction processes can be performed offline, thereby not impeding the real-time application of the models. In comparison, the classic Graph RAG approach relies on clustering to represent group correlations within the data, a process that is both time-consuming and resource-intensive. Light RAG, while alleviating the computational burden by omitting clustering, consequently loses high-order relational information, leading to diminished performance.

In the knowledge retrieval phase, Hyper-RAG offers distinct advantages over Graph-RAG by eliminating the need for redundant local and global retrieval processes. Instead, it allows for direct retrieval of nodes or relational structures. The retrieval of relational structures is optional; incorporating it can enhance performance but at the cost of additional computational resources. Alternatively, when only node information is retrieved, the system operates in a mode we designate as Hyper-RAG-Lite. In this mode, the integration of both low-order and high-order correlations enables the diffusion of information, thereby utilizing high-order knowledge embedded within the data. Consequently, Hyper-RAG-Lite not only accelerates the retrieval process but also improves the quality of responses generated by the LLMs, presenting a promising avenue for future research.

However, our current approach to knowledge construction faces challenges in directly extracting correlations across different data chunks, necessitating post-processing steps to merge these relations. A significant portion of the relevant information spans multiple chunks, making it inadequate to capture through a single chunk alone. Future research should focus on developing methods for the fusion and extraction of cross-chunk correlations. Additionally, modeling relationships between different documents could further enhance the dimensionality and scalability of the knowledge base. This study has demonstrated that improved representation and organization of domain knowledge can significantly enhance the capabilities of large language models. Future work may explore automating data organization and knowledge representation techniques, fostering a deeper integration with large language models to further mitigate issues such as hallucinations.

%% file: 3_method.tex
\section{Methods}
\label{sec:method}
In this section, we provide a comprehensive overview of the proposed Hypergraph-Driven Retrieval-Augmented Generation (Hyper-RAG) framework, encompassing the processes of knowledge extraction, indexing, and retrieval. 
Subsequently, we present the architecture of the open-domain question answering tasks utilized to evaluate Hyper-RAG, alongside two distinct evaluation metrics that assess both the accuracy and comprehensiveness of the generated responses. These metrics are chosen to rigorously measure the reduction in hallucinations and the enhancement in answer reliability provided by our approach.
Finally, we describe the data collection and construction procedures, detailing the sources and criteria for dataset assembly. Additionally, we illustrate the various prompt templates employed at different stages of the Hyper-RAG pipeline, demonstrating how they are tailored to optimize knowledge retrieval and generation processes. 

\begin{table}[!ht]
    \centering
    \caption{The comparison of different LLMs augmentation strategies.}
    \label{tab:cmp}
    \begin{tabularx}{\linewidth}{lXc}
        \toprule
         Method& Formulation & Reference \\ \midrule
         LLM & $\text{response} = \text{LLM}\left(\gP_q(q)\right)$  & \cite{gpt4,ds-v3,glm,qwen,llama,doubao}\\
         RAG & $\text{response} = \text{LLM}\left(\gP_q(q, \gD)\right)$ & \cite{es2024ragas,xiong2024benchmarking,naiverag,miao2024integrating,yu2024rankrag,yu2024evaluation,wang2024leave,prince2024opportunities}\\
         Graph-RAG & $\text{response} = \text{LLM}\left(\gP_q(q, \text{RAG}(q, \gD, \gV, \gE_\text{low}))\right)$ & \cite{graphrag,lightrag,zhu2024structugraphrag,jiang2024ragraph,he2024g}\\
         Hyper-RAG& $\text{response} = \text{LLM}\left(\gP_q(q, \text{RAG}(q, \gD, \gV, \gE_\text{low}, \gE_\text{high}))\right)$ & This work \\ 
         \bottomrule
    \end{tabularx}
\end{table}

\Cref{tab:cmp} presents a mathematical comparison of various LLMs enhancement strategies. Here, $\gP_q$ denotes the function that transforms the input into a prompt, where $q$ represents the input query and $\gD$ is the prior corpus. The symbols $\gV$, $\gE\text{low}$, and $\gE_\text{high}$ correspond to the vertices, low-order correlation knowledge, and high-order correlation knowledge of the knowledge base constructed from the corpus, respectively. It is evident that the original LLM generates responses directly based on the input question $q$. In contrast, RAG retrieves relevant data from the prior corpus $\gD$ to assist the LLMs in providing answers. Graph-RAG further extracts low-order structural information from the prior knowledge. Our proposed Hyper-RAG simultaneously constructs both low-order and high-order correlation information from the prior corpus, enabling a more comprehensive representation of knowledge. This enhanced knowledge representation effectively reduces information loss, thereby mitigating the occurrence of hallucinations in LLMs.





\subsection{Framework Schema}
\begin{figure}[!t]
    \centering
    \includegraphics[width=\linewidth]{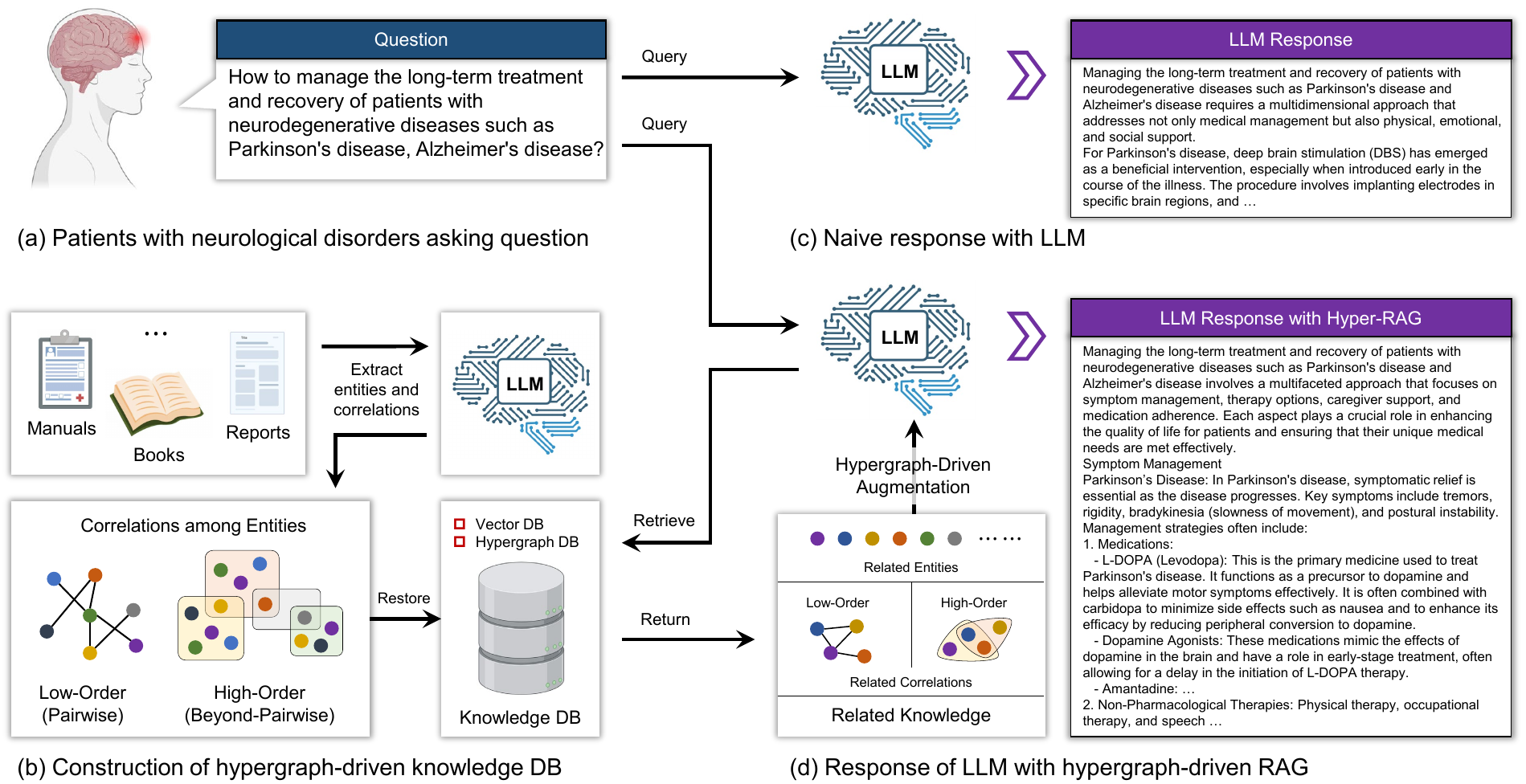}
    \caption{\textbf{Schematic diagram of the proposed Hyper-RAG architecture.} \textbf{a,} The patient poses a question. \textbf{b,} A knowledge base is constructed from relevant domain-specific corpora. \textbf{c,} Responses are generated directly using LLMs. \textbf{d,} Hyper-RAG generates responses by first retrieving relevant prior knowledge from the knowledge base and then inputting this knowledge, along with the patient's question, into the LLMs to formulate the reply.}
    \label{fig:fw}
\end{figure}

\Cref{fig:fw} illustrates the proposed Hyper-RAG framework, encompassing both the offline construction of the knowledge database and the online retrieval-augmented response generation processes. Initially, we collect domain-specific corpora, which may include manuals, books, reports, and other relevant documents. These raw corpora are then processed using LLMs to segment the text and extract entities and their relationships. In the Hyper-RAG framework, relationships are categorized into pairwise low-order correlations and beyond-pairwise high-order correlations that represent correlations among groups of entities. The extracted knowledge is subsequently stored in a database to facilitate rapid retrieval during the query phase.

During the question-answering process, consider an example where a user with neurological disorders poses a question to the LLMs, as shown in \cref{fig:fw}. When using a naive LLM, the model responds directly to the patient's query without additional context. In contrast, the Hyper-RAG strategy involves a two-step approach: first, we extract keywords from the user's question; second, we retrieve knowledge related to these similar keywords from the knowledge database. The retrieved relevant knowledge is then provided as supporting information alongside the original question to the LLMs, resulting in more accurate and reliable responses. The subsequent sections will provide a detailed description of each component and process within the Hyper-RAG.

\subsection{Knowledge Extraction}
The objective of knowledge extraction is to systematically organize raw corpora into a structured format, thereby enabling more efficient retrieval of prior information. In our approach, the corpus data can comprise various types of documents, including books, manuals, reports, and other relevant texts. We begin by preprocessing the original documents and partitioning them into uniformly sized chunks, denoted as $D_i$, thereby forming the corpus collection:
\begin{equation}
    \gD = \{D_1, D_2, \dots, D_N\}.
\end{equation}

Subsequently, a document structuring function, $\phi$, is employed to extract structural information from the corpus, resulting in a hypergraph $\mathcal{G}$:
\begin{equation}
    \gG = \phi(\gD) \quad \text{and} \quad \gG = \{\gV, \gE_\text{low}, \gE_\text{high}\},
\end{equation}
where $\mathcal{G}$ represents the hypergraph structure extracted from the documents, comprising a set of vertices $\gV$, low-order correlations $\gE_\text{low}$, and high-order correlations $\gE_\text{high}$. The elements within the vertex set $\gV$ can be of various types, such as names of entities, task titles, or skills. The relation sets $\gE_\text{low}$ and $\gE_\text{high}$ describe the connections between entities, where $\gE_\text{low}$ captures pairwise relationships and $\gE_\text{high}$ encapsulates correlations involving multiple entities.
For each chunk $D_i \in \gD$, we extract entities and their descriptions using LLMs as follows:
\begin{equation}
    \gK_v = \text{LLM}(\gP_\text{ext\_entity}(D_i)) \quad \text{for} \quad D_i \in \gD,
\end{equation}
where $\gK_v = \{v_1, v_2, \dots \}$ denotes the set of entities, each accompanied by a generated description, as illustrated in \cref{fig:extract}. It is important to note that if multiple chunks contain the same entity, their descriptions are merged using the LLMs to ensure consistency and completeness. The function $\gP_\text{ext\_entity}$ serves as the prompt filler that converts the input into an appropriate prompt for entity extraction, which is detailed in \cref{sec:prompts}.

Following entity extraction, we proceed to identify the corresponding low-order and high-order correlations within each chunk based on the extracted entities:
\begin{equation}
\left\{
\begin{aligned}
    \gK_e^\text{low} &= \text{LLM}(\gP_\text{ext\_low}(D_i, \gK_v))\\
    \gK_e^\text{high} &= \text{LLM}(\gP_\text{ext\_high}(D_i, \gK_v))
\end{aligned}
\right.  \quad \text{for} \quad D_i \in \gD,
\end{equation}
where $\gK_e^\text{low} = \{(u, v), \dots \}$ represents the set of low-order correlations between pairs of entities, while $\gK_e^\text{high} = \{(u, v, \dots), \dots \}$ denotes the high-order correlations involving multiple entities. Each relation within the knowledge base is also accompanied by a descriptive narrative, as depicted in \cref{fig:extract}. When identical correlations are extracted from different chunks, their descriptions are amalgamated to maintain a unified representation. This comprehensive extraction of both low-order and high-order relational information from the corpus ensures a robust and detailed knowledge base, which is critical for minimizing information loss and enhancing the retrieval process in the Hyper-RAG framework.

\subsection{Knowledge Indexing}
Hyper-RAG utilizes two distinct types of databases to effectively organize and manage the extracted knowledge: a vector database for storing the embedding representations of vertices and a hypergraph database for maintaining both high-order and low-order relational structures.

\paragraph{Vector Database}

The vector database stores fixed-dimensional vector representations derived from the descriptions of each entity. These embeddings are organized into a matrix $\mM$, where each row corresponds to the vector representation of an entity or a hyperedge. During retrieval, a query vector $\vq$ is compared against the vectors in the database using distance metrics such as cosine similarity or Euclidean distance. The system then retrieves the top-$k$ nearest vectors, which may correspond to relevant entities. This approach ensures that the most semantically similar entries are efficiently identified and utilized to enhance the generation process.

\paragraph{Hypergraph Database}

The hypergraph database stores the structural information extracted from the raw corpus, encompassing both low-order and high-order correlations. This database comprises two primary components: vertex adjacency lists and relation adjacency lists. A hypergraph extends a traditional graph by allowing hyperedges to connect more than two vertices, thereby uniformly storing both low-order (pairwise) and high-order (beyond-pairwise) correlations within this structure. Each hyperedge is represented as a tuple of vertex names, facilitating the representation of complex relationships among multiple entities. During retrieval, given a specific vertex name, the hypergraph database can swiftly access the vertex's descriptive information as well as its connected relational structures. Additionally, the database supports relational queries where, upon inputting a particular relation structure, it returns the corresponding descriptive information and the neighboring vertices associated with that relation. This dual capability allows Hyper-RAG to efficiently navigate and utilize intricate relational data, thereby enhancing the accuracy and contextual relevance of responses generated by the LLMs.

\paragraph{Integration of Vector and Hypergraph Databases}

The integration of the vector database and the hypergraph database within the Hyper-RAG framework provides a comprehensive mechanism for knowledge storage and retrieval. While the vector database excels in capturing and retrieving semantically similar entities and correlations through embedding spaces, the hypergraph database ensures that the structural and relational integrity of the knowledge base is maintained and easily accessible. Together, these databases enable Hyper-RAG to leverage both the semantic richness and the structural complexity of the underlying knowledge, thereby effectively mitigating hallucinations in large language models by providing accurate and contextually relevant information.

\subsection{Knowledge Retrieval and LLMs Augmentation}
After constructing the hypergraph knowledge base offline, we detail the methodology for augmenting the LLMs' response capabilities using knowledge databases. Given a user query $q$, we first extract two distinct sets of keywords: the entity keyword set $\mathcal{X}_\text{ent}$ (fundamental components) and the correlation keyword set $\mathcal{X}_\text{cor}$ (complex interdependencies), as follows:
\begin{equation}
    \gX_\text{ent}, \gX_\text{cor} = \text{LLM}(\gP_\text{ext\_key}(q)) \quad \text{and} \quad \gX_* = \{x_1, x_2, \dots \},
\end{equation}
where $\gP_\text{ext\_key}$ is the prompt used to extract keywords from the input question, as detailed in \cref{sec:prompts}. 
The entity keyword set comprises specific detailed nouns, such as personal names, locations, and other discrete identifiers. In contrast, the correlation keyword set encompasses more sophisticated descriptions, typically involving interactions between two or more entities. These correlations often capture narratives, systems, responses, reactions, and various forms of interactions that emerge from the relationships among entities. By distinguishing between these two types of keyword sets, our approach effectively models both the fundamental components and the complex interdependencies within the data. This comprehensive representation enhances the retrieval-augmented generation process, enabling the LLMs to leverage a richer and more nuanced foundation of prior knowledge.
Subsequently, based on these two categories of extracted keywords, we retrieve relevant information from the hypergraph database. It is important to note that entity keyword retrieval targets vertices, while correlation keyword retrieval targets hyperedges. This distinction arises because entity keywords predominantly describe individual entities, making vertices the appropriate retrieval objects. In contrast, correlation keywords describe abstract information that typically involves relationships among multiple entities, thereby necessitating hyperedges as retrieval targets.
For entity information retrieval, we employ the following formulation:
\begin{equation}
\left\{
\begin{aligned}
    \gV_\text{rel} &= \{\psi_\text{ret}(x_i, \gV) | x_i \in \gX_\text{ent} \} \quad // \text{Entity information} \\
    \gE_\text{more} &= \{e | v \in e \text{ and } v \in \gV_\text{rel} \} \quad // \text{Extended information via diffusion}
\end{aligned}
\right. ,
\end{equation}
where $\psi_\text{ret}$ denotes the vector-based retrieval function, which retrieves vertices similar to the input $x_i$ from the vertex vector database. Subsequently, $\gE_\text{more}$ is used to diffuse through the associated structural relationships, thereby obtaining hyperedges connected to these vertices as supplementary information.

Similarly, for correlation information retrieval, we use the following formulation:
\begin{equation}
\left\{
\begin{aligned}
    \gE_\text{rel} &= \{\psi_\text{ret}(x_i, \gE) | x_i \in \gX_\text{cor} \} \quad // \text{Correlation information}\\
    \gV_\text{more} &= \{v | v \in e \text{ and } e \in \gE_\text{rel} \} \quad // \text{Extended information via diffusion}
\end{aligned}
\right. ,
\end{equation}
where $\psi_\text{ret}(x_i, \gE)$ retrieves hyperedges related to the correlation keywords from the hyperedge vector database. Through one-step diffusion, the vertices associated with these hyperedges are acquired as supplementary information. Due to the constraints on the LLM's input context length, we aggregate and rank the retrieved entity and correlation information, selecting the most relevant data based on the maximum permissible context length to serve as prior knowledge for augmenting the LLM. In practical applications, we incorporate the textual content from the original chunks associated with the relevant vertices and hyperedges as prior information. This approach is employed to mitigate the potential hallucinations that may arise from descriptions synthesized by the LLM, thereby ensuring the reliability of the supplementary knowledge.

\subsection{Dataset}
To assess the efficacy of the proposed Hyper-RAG framework, we curate an extensive collection of corpora encompassing both domain-specific and mixed-domain datasets. Recognizing that domain-specific data is more susceptible to hallucinations, owing to the heightened demands for lexical precision in specialized fields, we selected nine corpora across eight distinct domains: medicine, mathematics, agriculture,  finance, physics, law, and art, the comprehensive statistics of which are detailed in \cref{tab:data}. Additionally, to evaluate the model’s performance in managing general knowledge across diverse areas, we construct a mixed-domain dataset. These corpora, referenced in \cite{xiong2024benchmarking,qian2024memorag}, are primarily extracted from books, reports, academic papers, narratives, and encyclopedias, with an average token count of $2,733,191$ per dataset. Each raw corpus underwent preprocessing to eliminate special symbols and non-textual elements, retaining solely the textual information. Subsequently, the sanitized corpora were partitioned into fixed-size chunks of 1200 tokens, with an overlapping segment of 100 tokens between consecutive chunks to ensure contextual coherence. For performance evaluation, a LLMs (GPT-4o mini) is employed to generate 50 questions per dataset. The question generation leverage the $\gP_\text{ext\_q}(q)$ prompt tailored to each chunk, facilitating automatic question formulation by the LLM. Furthermore, the origin chunk for each question are recorded to enable Scoring-Based Assessment, wherein the corresponding source chunk served as the reference answer for evaluating the responses.

\section{Evaluation Criteria}
\label{sec:metric}
The evaluation of LLMs has predominantly been conducted using benchmarks with predefined answers, such as SQuAD, GPT-3 Benchmarks, and others. These benchmarks typically involve generating concise responses that align with standard answers. However, in real-world applications, obtaining standard answers is often impractical, especially for open-ended questions where responses can vary widely. In such scenarios, the absence of supervisory information makes LLMs more prone to hallucinations, leading to the confusion of critical entities like names, dates, and locations. This issue is particularly detrimental in sensitive domains such as medicine, where inaccuracies can result in significant consequences. To effectively evaluate LLMs in open-ended contexts, we introduce two assessment strategies: Scoring-Based Assessment and Selection-Based Assessment.
\subsection{Scoring-Based Assessment}
\label{sec:score_metric}
Scoring-Based Assessment is designed to facilitate the comparative evaluation of multiple model outputs by quantifying their performance across various dimensions. This approach allows for a nuanced assessment of model capabilities by providing scores on several key metrics. However, a notable limitation is its reliance on reference answers. In our preprocessing steps, we leverage the source chunks from which each question is derived as reference answers. Using these references, we construct a scoring prompt, denoted as $\mathcal{P}_{\text{eval\_score}}$, which directs the LLM to evaluate open-ended responses based on five dimensions:

\begin{enumerate}
    \item \textbf{Comprehensiveness (0-100)}: Assesses whether the response sufficiently addresses all relevant aspects of the question without omitting critical information.
    \item \textbf{Diversity (0-100)}: Evaluates the richness of the content, including additional related knowledge beyond the direct answer.
    \item \textbf{Empowerment (0-100)}: Measures the credibility of the response, ensuring it is free from hallucinations and instills confidence in the reader regarding its accuracy.
    \item \textbf{Logical (0-100)}: Determines the coherence and clarity of the response, ensuring that the arguments are logically structured and well-articulated.
    \item \textbf{Readability (0-100)}: Examines the organization and formatting of the response, ensuring it is easy to read and understand.
\end{enumerate}

Each evaluation dimension is scored on a scale from 0 to 100, with higher scores indicating better performance. Recognizing the challenges associated with assigning broad numerical scores directly, we implemented a hierarchical scoring system by dividing each dimension into five distinct levels. Each level corresponds to specific criteria that provide clear and consistent guidelines for scoring. To illustrate, we present the classification for the \textbf{Comprehensiveness} dimension:

\begin{enumerate}
    \item \textbf{Level 1 | 0-20}: The answer is extremely one-sided, leaving out key parts or important aspects of the question.
    \item \textbf{Level 2 | 20-40}: The answer has some content but misses many important aspects and is not comprehensive enough.
    \item \textbf{Level 3 | 40-60}: The answer is more comprehensive, covering the main aspects of the question, but there are still some omissions.
    \item \textbf{Level 4 | 60-80}: The answer is comprehensive, covering most aspects of the question with few omissions.
    \item \textbf{Level 5 | 80-100}: The answer is extremely comprehensive, covering all aspects of the question with no omissions, enabling the reader to gain a complete understanding.
\end{enumerate}

While the detailed five-level classification is exemplified for the \textbf{Comprehensiveness} dimension, similar hierarchical structures have been established for the other evaluation metrics (\textbf{Diversity}, \textbf{Empowerment}, \textbf{Logical}, and \textbf{Readability}) to ensure uniformity and precision in the scoring process. Finally, an overall performance score is calculated as the average of the individual dimension scores. A higher overall performance score indicates greater accuracy in expression and a lower probability of hallucinations.

\subsection{Selection-Based Assessment}
\label{sec:selection_metric}
Selection-Based Assessment is tailored for scenarios where preliminary candidate models are available, enabling a comparative evaluation through a binary choice mechanism. This method does not require reference answers, making it suitable for diverse and open-ended questions. However, its limitation lies in its comparative nature, as it only allows for the evaluation of two models at a time.

In this strategy, the outputs from two methods, denoted as $ A_{\text{out}} $ and $ B_{\text{out}} $, are simultaneously presented to the LLM, denoted as $\mathcal{P}_{\text{eval\_select}}$. The model is then instructed to select the better response based on eight evaluation criteria:

\begin{enumerate}
    \item \textbf{Comprehensiveness}: How much detail does the answer provide to cover all aspects and details of the question?
    \item \textbf{Empowerment}: How well does the answer help the reader understand and make informed judgments about the topic?
    \item \textbf{Accuracy}: How well does the answer align with factual truth and avoid hallucination based on the retrieved context?
    \item \textbf{Relevance}: How precisely does the answer address the core aspects of the question without including unnecessary information?
    \item \textbf{Coherence}: How well does the system integrate and synthesize information from multiple sources into a logically flowing response?
    \item \textbf{Clarity}: How well does the system provide complete information while avoiding unnecessary verbosity and redundancy?
    \item \textbf{Logical}: How well does the system maintain consistent logical arguments without contradicting itself across the response?
    \item \textbf{Flexibility}: How well does the system handle various question formats, tones, and levels of complexity?
\end{enumerate}

For each of these eight criteria, the LLM selects the superior response between $ A_{\text{out}} $ and $ B_{\text{out}} $. The cumulative votes across all criteria determine the overall score for each model. This voting mechanism ensures a balanced evaluation based on multiple facets of response quality, thereby providing a robust assessment of the models' relative performance without the need for predefined reference answers.

\clearpage

\section{Prompts}
\label{sec:prompts}
\subsection{Extracting Entities, Correlations and Keywords}

\begin{myprompt}{Extracting Entities}
    \textbf{Formulation}: $\gP_\text{ext\_entity}(D_i)$\newline 
    $D_i$ denotes the text chunk. \newline\newline
    \textbf{Prompt: }\textit{
    Identify all entities. For each identified entity, extract the following information:\newline
    - entity\_name: Name of the entity, use same language as input text. If English, capitalized the name.\newline
    - entity\_type: One of the following types: [{entity\_types}]\newline
    - entity\_description: Comprehensive description of the entity's attributes and activities.\newline
    - additional\_properties: Other attributes possibly associated with the entity, like time, space, emotion, motivation, etc.
}
\end{myprompt}

\begin{myprompt}{Extracting Low-Order Correlations}
    \textbf{Formulation}: $\gP_\text{ext\_low}(D_i, \gK_v)$\newline 
    $D_i$ denotes the text chunk, $\gK_v$ denotes the extracted entities. \newline\newline
    \textbf{Prompt: }\textit{
    From the entities identified in \{$\gK_v$\}, identify all pairs of (source\_entity, target\_entity) that are *clearly related* to each other.\newline
    For each pair of related entities, extract the following information:\newline
    - entities\_pair: The name of source entity and target entity, as identified in \{$\gK_v$\}.\newline
    - low\_order\_relationship\_description: Explanation as to why you think the source entity and the target entity are related to each other.\newline
    - low\_order\_relationship\_keywords: Keywords that summarize the overarching nature of the relationship, focusing on concepts or themes rather than specific details.\newline
    - low\_order\_relationship\_strength: A numerical score indicating the strength of the relationship between the entities.)
}
\end{myprompt}

\begin{myprompt}{Extracting High-Order Correlations}
    \textbf{Formulation}: $\gP_\text{ext\_high}(D_i, \gK_v)$\newline 
    $D_i$ denotes the text chunk, $\gK_v$ denotes the extracted entities. \newline\newline
    \textbf{Prompt: }\textit{
    Extract high-level keywords that summarize the main idea, major concept, or themes of the important passage.\newline
    (Note: The content of high-level keywords should capture the overarching ideas present in the document, avoiding vague or empty terms).
    \newline\newline
    For the entities identified in $\gK_v$, based on the entity pair relationships and the high-level keywords, find connections or commonalities among multiple entities and construct high-order associated entity set as much as possible.\newline
    (Note: Avoid forcibly merging everything into a single association. If high-level keywords are not strongly associated, construct separate association).\newline
    Extract the following information from all related entities, entity pairs, and high-level keywords:\newline
    - entities\_set: The collection of names for elements in high-order associated entity set, as identified in $\gK_v$.\newline
    - high\_order\_relationship\_description: Use the relationships among the entities in the set to create a detailed, smooth, and comprehensive description that covers all entities in the set, without leaving out any relevant information.\newline
    - high\_order\_relationship\_generalization: Summarize the content of the entity set as concisely as possible.\newline
    - high\_order\_relationship\_keywords: Keywords that summarize the overarching nature of the high-order association, focusing on concepts or themes rather than specific details.\newline
    - high\_order\_relationship\_strength: A numerical score indicating the strength of the association among the entities in the set.
}
\end{myprompt}

\begin{myprompt}{Extracting Keys from User Query}
    \textbf{Formulation}: $\gP_\text{ext\_key}(q)$\newline 
    $q$ denotes user input. \newline\newline
    \textbf{Prompt: }\textit{
    You are a helpful assistant tasked with identifying both high-level and low-level keywords in the user's query.\newline
    ---Goal---\newline
    Given the query, list both high-level and low-level keywords. High-level keywords focus on overarching concepts or themes, while low-level keywords focus on specific entities, details, or concrete terms.
}
\end{myprompt}

\subsection{Evaluation}
\begin{myprompt}{Evaluation of Scoring-Based Assessment}
    \textbf{Formulation}: $\gP_\text{eval\_scoring}(q, R, T_o)$\newline 
    $q$ denotes user input, $R$ denotes LLM response, $T_o$ denotes the original text chunk that generated the question. \newline\newline
    \textbf{Prompt: }\textit{You are an expert tasked with evaluating answers to the questions by using the relevant documents based on five criteria: Comprehensiveness, Diversity, Empowerment, Logical, and Readability.\newline\newline
    ---Goal---\newline
     You will evaluate tht answers to the questions by using the relevant documents based on five criteria:Comprehensiveness, Diversity, Empowerment, Logical, and Readability.\newline\newline
    -Comprehensiveness-\newline
    Measure whether the answer comprehensively covers all key aspects of the question and whether there are omissions.\newline
    Level   $|$ score range $|$ description\newline
    Level 1 $|$ 0-20   $|$ The answer is extremely one-sided, leaving out key parts or important aspects of the question.\newline
    Level 2 $|$ 20-40  $|$ The answer has some content, but it misses many important aspects of the question and is not comprehensive enough.\newline
    Level 3 $|$ 40-60  $|$ The answer is more comprehensive, covering the main aspects of the question, but there are still some omissions.\newline
    Level 4 $|$ 60-80  $|$ The answer is comprehensive, covering most aspects of the question, with few omissions.\newline
    Level 5 $|$ 80-100 $|$ The answer is extremely comprehensive, covering all aspects of the question with no omissions, enabling the reader to gain a complete understanding.\newline
    ...,\newline
    For each indicator, please give the problem a corresponding Level based on the description of the indicator, and then give a score according to the score range of the level.\newline\newline
    Here are the question: $q$\newline
    Here are the relevant document: $T_o$\newline
    Here are the answer: $R$\newline\newline
    Evaluate all the answers using the five criteria listed above, for each criterion, provide a summary description, give a Level based on the description of the indicator, and then give a score based on the score range of the level.
}
\end{myprompt}

\begin{myprompt}{Evaluation of Selection-Based Assessment}
    \textbf{Formulation}: $\gP_\text{eval\_scoring}(q, R_a, R_b)$\newline 
    $q$ denotes user input, $R_a$ denotes the response from one LLMs, $R_b$ denotes the response from another LLMs. \newline\newline
    \textbf{Prompt: }\textit{You will evaluate two answers to the same question based on eight criteria: Comprehensiveness, Empowerment, Accuracy, Relevance, Coherence, Clarity, Logical, and Flexibility.\newline\newline
    ---Goal---\newline
    You will evaluate two answers to the same question by using the relevant documents based on eight criteria: Comprehensiveness, Empowerment, Accuracy, Relevance, Coherence, Clarity, Logical, and Flexibility.\newline\newline
    -Comprehensiveness: How much detail does the answer provide to cover all aspects and details of the question?\newline
    -Empowerment: How well does the answer help the reader understand and make informed judgments about the topic?\newline
    ...,\newline
    -Flexibility: How well does the system handle various question formats, tones, and levels of complexity?\newline\newline
    For each criterion, choose the better answer (either Answer 1 or Answer 2) and explain why. Then, select an overall winner based on these ten categories.\newline\newline
    Here are the question: $q$\newline
    Here are the two answers: \newline
    Answer 1: $R_a$;\newline
    Answer 2: $R_b$\newline\newline
    Evaluate both answers using the eight criteria listed above and provide detailed explanations for each criterion.
}
\end{myprompt}